\documentstyle[emulateapj]{article}

\lefthead{Nagao et al.}
\righthead{Metallicity of NLRs in NLS1s and BLS1s}

\begin{document}

\submitted{The Astrophysical Journal, in Press} 
\title{Gas Metallicity of Narrow-Line Regions in
       Narrow-Line Seyfert 1 Galaxies and Broad-Line Seyfert 1 Galaxies}

\author{Tohru NAGAO, Takashi MURAYAMA, Yasuhiro SHIOYA, and Yoshiaki TANIGUCHI}
\affil{Astronomical Institute, Graduate School of Science, 
       Tohoku University, Aramaki, Aoba, Sendai 980-8578, Japan\\
       tohru@astr.tohoku.ac.jp, murayama@astr.tohoku.ac.jp, 
       shioya@astr.tohoku.ac.jp, tani@astr.tohoku.ac.jp}


\begin{abstract}

We investigate gas metallicity of narrow-line regions in
narrow-line Seyfert 1 galaxies (NLS1s) and broad-line ones (BLS1s) 
in order to examine whether or not there is a
difference in the gas metallicity between the two
populations of Seyfert 1 galaxies.
We apply two methods to study this issue.
One is to use the emission-line flux ratio of 
[N {\sc ii}]$\lambda$6583/H$\alpha_{\rm narrow}$ in combination with
some other optical emission-line flux ratios.
This method, which has been often applied to Seyfert 2 galaxies,
suggests that the gas metallicity of narrow-line regions is 
indistinguishable or possibly higher in BLS1s than in NLS1s.
On the contrary, the other method in which only forbidden emission-line 
fluxes are used results in that NLS1s tend to possess 
metal-richer gas in the narrow-line regions than BLS1s.
We point out that this inconsistency
may be owing to the contamination of the broad component of permitted
lines into the narrow component of ones in the first method.
Since the results derived by using only forbidden emission-line fluxes
do not suffer from any uncertainty of the fitting function for the broad 
component of Balmer lines, the results from this method are more reliable 
than those derived by using permitted lines.
We thus conclude that the gas metallicity of narrow-line regions tends to 
be higher in NLS1s than in BLS1s.

\end{abstract}



\section{INTRODUCTION}

\begin{deluxetable}{lccccc}
\tablenum{1}
\tablecaption{Statistical Properties of Narrow Emission-Line 
              Flux Ratios\tablenotemark{a}
              \hspace{0.1mm} of the NLS1s and the BLS1s \label{tbl-1}}
\tablehead{
\colhead{Line Ratio} &
\multicolumn{2}{c}{NLS1\tablenotemark{b}} &
\multicolumn{2}{c}{BLS1\tablenotemark{b}} &
\colhead{$P_{\rm KS}$} \\
\colhead{} &
\colhead{average} &
\colhead{median} &
\colhead{average} &
\colhead{median} &
\colhead{} 
}
\startdata
[N {\sc ii}]$\lambda$6583/H$\alpha$ 
   & 0.406 & 0.234 & 0.462 & 0.367 & 0.912 \nl
[O {\sc i}]$\lambda$6300/[N {\sc ii}]$\lambda$6583
   & 0.123 & 0.100 & 0.261 & 0.204 & 0.133 \nl
[O {\sc ii}]$\lambda$3727/[N {\sc ii}]$\lambda$6583
   & 0.281 & 0.237 & 0.686 & 0.489 & 0.533 \nl
[S {\sc ii}]$\lambda \lambda$6717,6731/[N {\sc ii}]$\lambda$6583
   & 0.427 & 0.423 & 0.590 & 0.433 & 0.767
\enddata
\tablenotetext{a}{Dust extinction is not corrected.}
\tablenotetext{b}{Data are taken from Rodr\'{\i}guez-Ardila et al. (2000)}
\end{deluxetable}

\begin{deluxetable}{lcccc}
\tablenum{2}
\tablecaption{Statistical Properties of the Estimated Oxygen Abundances
              of NLRs in the NLS1s and BLS1s \label{tbl-2}}
\tablehead{
\colhead{} &
\multicolumn{2}{c}{NLS1} &
\multicolumn{2}{c}{BLS1}\\
\colhead{} &
\colhead{average} &
\colhead{median} &
\colhead{average} &
\colhead{median}
}
\startdata
12 + log (O/H)$_{\rm SSCK1}\tablenotemark{a}$ 
   & 8.420 & 8.384 & 8.517 & 8.512 \nl
12 + log (O/H)$_{\rm SSCK2}\tablenotemark{a}$ 
   & 8.679 & 8.739 & 8.654 & 8.752 
\enddata
\tablenotetext{a}{Calculated by equations (1) and (2), which are derived
                  by Storchi-Bergmann et al. (1998).}
\end{deluxetable}

Seyfert nuclei are typical active galactic nuclei (AGNs) in the nearby 
Universe. They have been broadly classified into two types based on the 
presence or absence of broad ($\gtrsim$ 2000 km$^{-1}$) permitted lines 
in their optical spectra (Khachikian \& Weedman 1974); 
Seyfert nuclei with broad lines are type 1
(hereafter S1) while those without broad lines are type 2 (S2).
This difference can be explained by the ``unified model'' in which it 
is considered that the broad-line region (BLR) is surrounded by a dusty 
torus and that the BLR in S2s is obscured by the edge-on view torus from 
the observer (e.g., Antonucci \& Miller 1985; Antonucci 1993).
Therefore only narrow ($\lesssim$ 2000 km$^{-1}$) lines, which arise
from narrow-line regions (NLRs), are seen in the spectra of S2s.
In addition to these two types of Seyfert nuclei, narrow-line Seyfert 1
galaxies (NLS1s) also comprise a sub-class of the Seyfert nuclei.
Since NLS1s exhibit only narrow (i.e., $\lesssim$ 2000 km$^{-1}$)
permitted lines in their optical spectra, NLS1s are recognized as a
distinct type of ordinary broad-line S1s (BLS1s). However, NLS1s also 
show several characteristics of S1s (e.g., Osterbrock \& Pogge 1985);
e.g., small line flux ratios of
[O {\sc iii}]$\lambda$5007/H$\beta$ ($\lesssim$ 3), strong Fe {\sc ii}
and some high-ionization emission lines, and non-absorbed X-ray spectra.
Therefore, it is believed that the BLR of NLS1s is not obscured but
the line width of the BLR is narrow for some reason.

One possible idea to explain the origin of the narrow BLR emission of 
NLS1s is that NLS1s have a smaller mass black hole than BLS1s (e.g., 
Boller, Brandt, \& Fink 1996; Laor et al. 1997).
If the BLR in NLS1s is the same as the BLR in BLS1s but the gas motion
is more quiescent because of the smaller black hole mass, the narrowness 
of the BLR emission can be successfully explained.
The steep and highly variable X-ray spectra, which are also the common
properties of NLS1s (e.g., Boller et al. 1996; Turner et al. 1999a;
Leighly 1999a, 1999b), can be also
explained by introducing the small black hole mass (e.g., 
Wang, Brinkmann, \& Bergeron 1996; Hayashida 2000;
Mineshige et al. 2000; Lu \& Yu 2001; Puchnarewicz et al. 2001).

Recently, it has been proposed that NLS1s may be relatively young AGNs
with a black hole still in a growing-up phase (e.g., 
Mathur 2000a, 2000b; Mathur, Kuraszkiewicz, \& Czerny 2001;
see also Komossa \& Mathur 2001; Mathur 2001).
If NLS1s possess a smaller mass black hole than BLS1s while the luminosity
is similar between NLS1s and BLS1s, a mass accretion rate for the same black
hole mass is larger in NLS1s than in BLS1s. This implies that the mass
accretion in NLS1s is more efficient than in BLS1s, which leads to the 
above hypothesis. Indeed some observations suggest that the mass 
accretion rate in NLS1s is high, close to the Eddington rate 
(e.g., Pounds, Done, \& Osborne 1995; Kuraszkiewicz et al. 2000;
Pounds \& Vaughan 2000; Wandel 2000; Puchnarewicz et al. 2001).
Moreover, Wandel (2002) reported that NLS1s have the smaller black hole
mass than the value inferred from the correlation between the black hole mass
and spheroidal mass of the host galaxy (see also Wandel 1999, 2000;
Mathur et al. 2001), which is consistent with the hypothesis that 
NLS1s possess a growing black hole in their nucleus.

In the framework of this scheme, the gas metallicity of NLS1s has been
discussed very recently. 
Wills et al. (1999) found a negative correlation between 
the emission-line width of H$\beta$ and the line flux ratio of
N {\sc v} $\lambda$1240/C {\sc iii}]$\lambda$1909, which can be interpreted
as a result of a nitrogen overabundance in BLRs of NLS1s
(see also Wills, Shang, \& Yuan 2000; Shemmer \& Netzer 2002).
The observed strong fluorescent Fe K$\alpha$ and optical Fe {\sc ii} 
emission, and the unusual spectral feature at 1 keV, which are seen in
spectra of some NLS1s, may suggest the overabundance of iron in
NLS1s (Turner, George, \& Netzer 1999; Ulrich et al. 1999;
Collin \& Joly 2000; see also Boller et al. 2002).
In order to investigate the gas metallicity of NLS1s and BLS1s further, 
it seems crucial to examine
the gas metallicity in NLS1s via various approaches.
Therefore, in this paper, we study the gas metallicity of NLRs in NLS1s and
compare it with that in BLS1s using some optical
diagnostic emission-line flux ratios.

\section{DATA}

\begin{figure*}
\epsscale{0.60}
\plotone{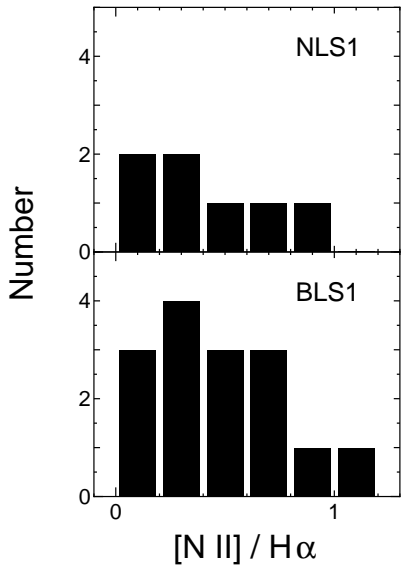}
\caption{
Frequency distributions of the emission-line flux ratio of
[N {\sc ii}]$\lambda$6583/H$\alpha$ for the NLS1s and the BLS1s
in the sample of Rodr\'{\i}guez-Ardila et al. (2000).
The data are not corrected for the dust extinction.
\label{fig1}}
\end{figure*}

The gas metallicity in NLRs of AGNs has been often investigated for S2s 
(e.g., Shields \& Oke 1975; Storchi-Bergmann \& Pastoriza 1989, 1990; 
Storchi-Bergmann 1991; Alloin et al. 1992; Kraemer et al. 1994;
Schmitt, Storchi-Bergmann, \& Baldwin 1994; Radovich \& Rafanelli 1996;
Kraemer, Ruiz, \& Crenshaw 1998) while it has been scarcely studied for 
S1s (e.g., Kraemer et al. 1999).
One reason for this is that the accurate measurement of the
narrow component of permitted lines is rather difficult for S1s.
Thus high-quality spectra and their careful analysis are required to
measure the narrow-component flux of permitted lines.

Recently, Rodr\'{\i}guez-Ardila, Pastoriza, \& Donzelli (2000) presented
high-quality optical spectra of 7 NLS1s\footnote{
   CTS H34.06, Mrk 1239, CTS J03.19, CTS J04.08, NGC 4748, CTS J13.12,
   and 1H 1934--063.} 
and 16 BLS1s\footnote{
   CTS C16.16, MCG --5-13-17, CTS H34.03, CTS B31.01, Fairall 1146,
   CTS M02.30, CTS J07.02, CTS J10.09, CTS R12.15, CTS J14.05,
   CTS J15.22, CTS M17.17, CTS G03.04, 1H 2107--097, CTS A08.12,
   and CTS F10.01. Note that we do not use the data of CTS A08.12 in the
   following analysis since the flux of the [N {\sc ii}]$\lambda$6583 
   emission of this object is not presented in 
   Rodr\'{\i}guez-Ardila et al. (2000).}.
They carefully removed the underlying stellar continuum and multiple
Fe {\sc ii} emission lines from the observed spectra.
Then they deconvolved the narrow component of permitted lines from the broad
component and measured various emission-line fluxes accurately.
We basically refer the data \footnote{
   Note that there is a type-setting error in the Rodr\'{\i}guez-Ardila et al. 
   (2000) paper; i.e., a second header of Table 4 in the 
   Rodr\'{\i}guez-Ardila et al. (2000) paper was cut in the editing process. 
   The second set of line fluxes (when the object names are repeated) 
   corresponds in fact to other emission lines organized as follows:
   [O {\sc i}]$\lambda$6300, [Fe {\sc x}]$\lambda$6374, 
   H $\alpha \lambda$6563, [N {\sc ii}]$\lambda$6584, 
   [S {\sc ii}]$\lambda$6717, [S {\sc ii}]$\lambda$6731, 
   [Ar {\sc iii}]$\lambda$7135, [O {\sc ii}]$\lambda$7325, 
   [S {\sc iii}]$\lambda$9069, and $E$($B-V$).}
of Rodr\'{\i}guez-Ardila et al. (2000) in this paper to investigate 
the gas metallicity of NLRs in NLS1s and BLS1s.

In order to see some possible correlations among narrow emission-line 
flux ratios in a larger sample of Seyfert nuclei, we also refer the 
data compiled by Nagao, Murayama, \& Taniguchi (2001c). Their sample 
contains $\approx$350 Seyfert nuclei.
Since their work is based on the data compilation from the literature,
the measurement of emission-line flux ratios is done in heterogeneous ways.
Therefore, we refer their results only when we see general trends of 
emission-line flux ratios and we never use them to discuss the 
difference in the properties of the NLR emission between NLS1s and BLS1s.

\section{PREVIOUS DIAGNOSTICS FOR THE GAS METALLICITY OF NLR}

 \subsection{The [N {\sc ii}]$\lambda$6583/H$\alpha$ Method}

The gas metallicity in NLRs of S2s has been sometimes studied by using the
emission-line flux ratio of [N {\sc ii}]$\lambda$6583/H$\alpha$
(e.g., Storchi-Bergmann \& Pastoriza 1989, 1990; Storchi-Bergmann 1991;
Radovich \& Rafanelli 1996).
Using photoionization models, Storchi-Bergmann \& Pastoriza (1989, 1990) 
showed that this emission-line flux ratio is sensitive to the nitrogen 
abundance but rather insensitive to the gas density and the ionization
parameter (i.e., a number density ratio of ionizing photons to hydrogen 
atoms). Note that the nitrogen abundance is important since the N/O 
abundance ratio in galactic H {\sc ii} regions is known to scale with 
the O/H abundance ratio (e.g., Shields 1976; Pagel \& Edmunds 1981; 
Villa-Costas \& Edmunds 1993; van Zee, Salzer \& Haynes 1998; 
Izotov \& Thuan 1999). Therefore the nitrogen abundance, N/H, scales
with $Z^2$, where $Z$ is the gas metallicity. The data of H {\sc ii} regions 
show that these scaling relations hold approximately for 
$Z \gtrsim 0.2 Z_{\odot}$.

We show the histograms of the emission-line flux ratio of 
[N {\sc ii}]$\lambda$6583/H$\alpha_{\rm narrow}$ for the NLS1s and the 
BLS1s in the sample of Rodr\'{\i}guez-Ardila et al. (2000) in Figure 1. 
The data are not corrected for the dust extinction
because the effect of the dust extinction on this flux ratio is 
negligibly small. The average and median values of 
[N {\sc ii}]$\lambda$6583/H$\alpha_{\rm narrow}$ 
for the NLS1s and those for the BLS1s are given in Table 1.
There is no apparent difference in this flux ratio between the two 
populations.
It is also suggested by the Kolmogorov-Smirnov (KS) statistical test that
the two frequency distributions of the flux ratio of
[N {\sc ii}]$\lambda$6583/H$\alpha_{\rm narrow}$ are statistically 
indistinguishable ($P_{\rm KS} = 0.912$). As shown in Figure 2,
we cannot find an apparent correlation between the flux ratio of
[N {\sc ii}]$\lambda$6583/H$\alpha_{\rm narrow}$ and full-width at 
half maximum (FWHM) of the broad component of the H$\alpha$ emission.
The corresponding correlation coefficient is 0.009, which implies that
there is no meaningful correlation between the two quantities.
This result seems contrary to the result of Wills et al. (1999),
who found an apparent negative correlation between the emission-line flux 
ratio of N {\sc v} $\lambda$1240/C {\sc iii}]$\lambda$1909, which is sensitive
to the nitrogen abundance, and the FWHM of the broad component of the 
H$\beta$ emission.

 \subsection{The SSCK Method}

\begin{figure*}
\epsscale{1.00}
\plotone{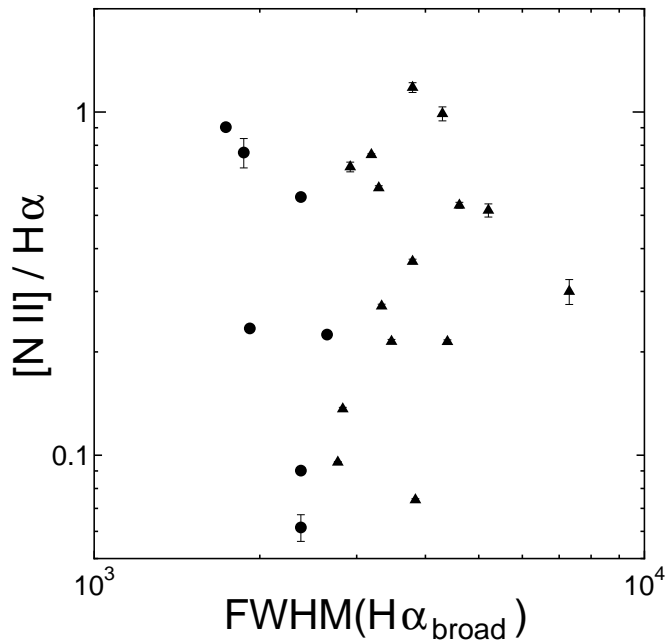}
\caption{
Diagram of the flux ratio of [N {\sc ii}]$\lambda$6583/H$\alpha$ versus
the FWHM of the broad component of the H$\alpha$ emission.
The circles and the triangles denote the NLS1s and the BLS1s of
the sample of Rodr\'{\i}guez-Ardila et al. (2000), respectively.
\label{fig2}}
\end{figure*}

Based on the intensive photoionization model calculations and the
careful calibrations with observations, Storchi-Bergmann et al. 
(1998, hereafter SSCK) proposed
the revised diagnostics for the gas metallicity of NLRs in AGNs.
They showed that the oxygen abundance of the NLR gas in AGNs can be
expressed by
\begin{eqnarray}
{\rm 12 + log (O/H)_{SSCK1}} &=& 
8.34 + 0.212x - 0.012x^2 \nonumber \\
& & {} - 0.002y + 0.007xy - 0.002x^2 y \nonumber \\
& & {} + 6.52 \times 10^{-4} y^2 \nonumber \\
& & {} + 2.27 \times 10^{-4} xy^2  \nonumber \\
& & {} + 8.87 \times 10^{-5} x^2 y^2
\end{eqnarray}
where $x \equiv$ [N {\sc ii}]$\lambda$6583/H$\alpha_{\rm narrow}$ and 
$y \equiv$ [O {\sc iii}]$\lambda$5007/H$\beta_{\rm narrow}$, and
\begin{eqnarray}
{\rm 12 + log (O/H)_{SSCK2}} &=& 
8.643 - 0.275u + 0.164u^2 \nonumber \\
& & {} + 0.655v - 0.154uv \nonumber \\
& & {} - 0.021 u^2 v + 0.288v^2 \nonumber \\
& & {} + 0.162uv^2 + 0.0353u^2 v^2
\end{eqnarray}
where $u \equiv$ log ([O {\sc ii}]$\lambda$3727/[O {\sc iii}]$\lambda$5007)
and $v \equiv$ log ([N {\sc ii}]$\lambda$6583/H$\alpha_{\rm narrow}$).
These methods weakly depend on the gas density; the derived values should
be subtracted by the correction term, 
0.1 $\times$ log ($n_{\rm H}$/300 cm$^{-3}$) where $n_{\rm H}$ is the
hydrogen density (see SSCK for more details).
We apply these relations to the data of Rodr\'{\i}guez-Ardila et al. (2000).
For the objects whose gas density is not given in 
Rodr\'{\i}guez-Ardila et al. (2000) (CTS H34.03, CTS M02.30, CTS J14.05,
CTS G03.04, and 1H 2107-097), we assume the gas density to be 300 cm$^{-3}$.
The average and median values of the derived 12 + log (O/H)$_{\rm SSCK1}$
and 12 + log (O/H)$_{\rm SSCK2}$ are given in Table 2.
Here the effect of the dust extinction is corrected by adopting the values
of $A_V$ presented by Rodr\'{\i}guez-Ardila et al. (2000) and the
extinction curve of Cardelli, Clayton, \& Mathis (1989).
There is no tendency that the NLS1s are metal rich compared to the BLS1
as implied by some earlier studies. To the contrary, the derived oxygen 
abundances may be larger in the sample of BLS1s than in the sample of NLS1s.
We show the relationships between the FWHM of the broad component of the 
H$\alpha$ emission and the derived oxygen abundances in Figure 3.
The object with a larger FWHM of the H$\alpha$ emission seems to be
more oxygen abundant when equation (1) is applied, though this 
tendency is not significant. On the other hand,
such a tendency is not seen when equation (2) is applied.

 \subsection{Remarks on the Two Methods}

\begin{figure*}
\epsscale{0.80}
\plotone{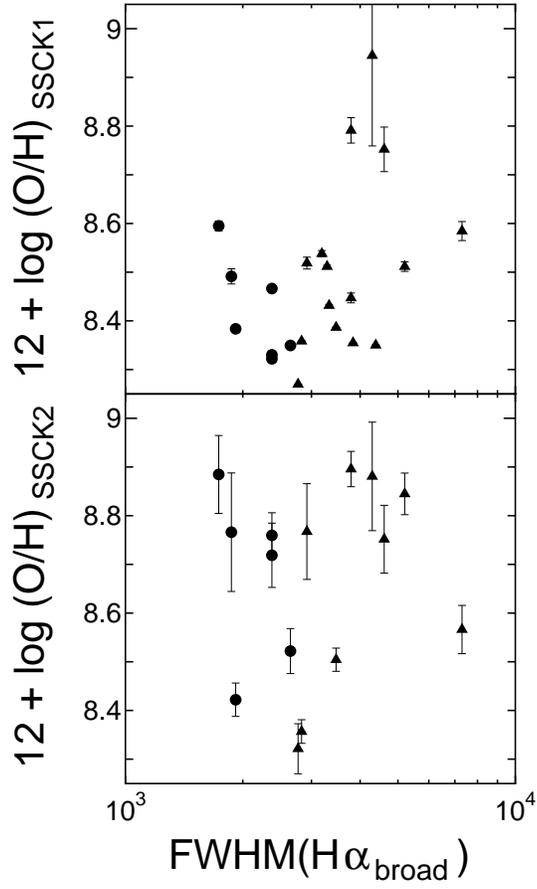}
\caption{
Diagram of the oxygen abundances estimated by the equations given by
Storchi-Bergmann et al. (1998) versus the FWHM of the broad component of 
the H$\alpha$ emission.
The symbols are the same as those in Figure 2.
\label{fig3}}
\end{figure*}

\begin{figure*}
\epsscale{1.00}
\plotone{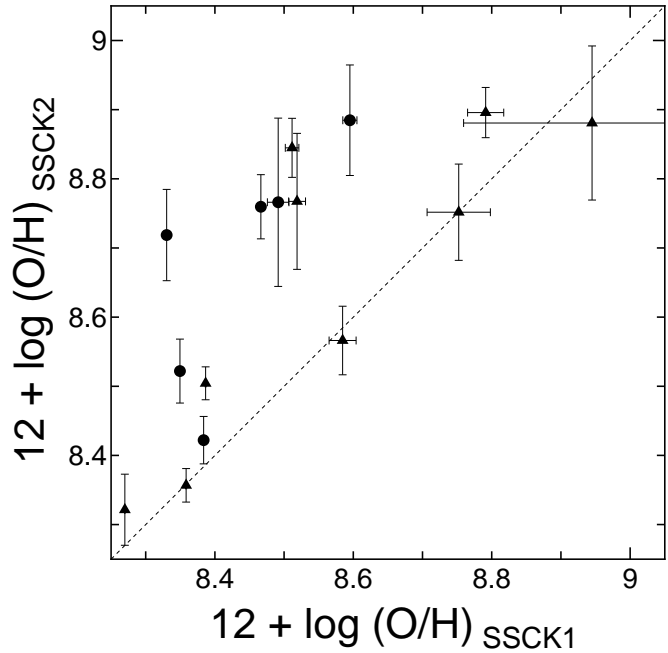}
\caption{
Diagram of the oxygen abundance estimated by the equation (1)
versus that by the equation (2).
The loci of equal values for the two quantities are drawn by a dot line
for comparison.
The symbols are the same as those in Figure 2.
\label{fig4}}
\end{figure*}

Do the above results presented in \S 3.2 suggest that 
the NLR of NLS1s is not metal rich compared to that of BLS1s?
Recently, V\'{e}ron-Cetty, V\'{e}ron, \& Gon\c calves (2001) pointed out 
that the broad component of permitted lines in the optical spectra of NLS1s
is well fitted by a Lorentzian 
profile and thus the decomposition of the narrow component of permitted 
lines from the broad component by Rodr\'{\i}guez-Ardila et al. (2000) may 
be inaccurate, since the broad components of permitted lines were fitted by
a Gaussian profile by Rodr\'{\i}guez-Ardila et al. (2000) 
(see also Moran, Halpern, \& Helfand 1996; Leighly et al. 1999b;
Gon\c calves, V\'{e}ron, \& V\'{e}ron-Cetty 1999; Sulentic et al. 2002). 
If this claim is the case,
the flux ratio of [N {\sc ii}]$\lambda$6583/H$\alpha_{\rm narrow}$ may be
inappropriate for the metallicity diagnostics because the selection of the
fitting function for the broad component of permitted lines affects
the measurement of not only the broad component but also the
narrow component of permitted lines. 

Indeed, the derived oxygen abundances of the NLR gas of the NLS1s and 
the BLS1s by using equations (1) and (2) should be suspected if 
we take account of the following two facts. 
First, most of the derived oxygen abundances suggest 
very sub-solar metallicities.
This is especially remarkable when equation (1) is adopted:
the median value of the derived oxygen abundances corresponds to
$Z \simeq 0.32 Z_{\odot}$.
Second, the derived oxygen abundances are not consistent to
each other. That is, the average and median values of the derived 
oxygen abundances are systematically different: the oxygen abundances 
derived by using equation (2) are larger than that by using
equation (1) by $\sim$0.3 dex. This difference is more clearly shown in
Figure 4, in which we compare the two kinds of the inferred
oxygen abundances.
Although SSCK also mentioned that the oxygen abundance derived by
equation (2) tends to be higher than the one derived by
equation (1) by $\sim$0.11 dex, the difference in the sample of
Rodr\'{\i}guez-Ardila et al. (2000) is much larger than that
reported by SSCK. This may imply that the methods proposed by SSCK
are inappropriate to estimate the gas metallicities of NLRs, 
at least for S1s. 
Thus it seems to be safe to avoid using the
narrow component of permitted lines for the investigation of the
gas properties of NLRs in S1s.

\begin{figure*}
\epsscale{0.60}
\plotone{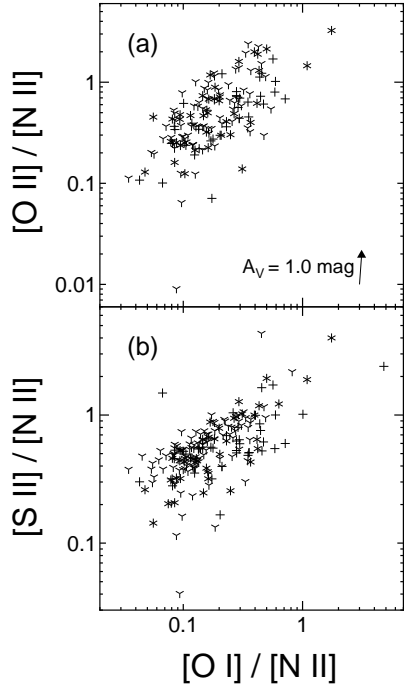}
\caption{
Diagrams of emission-line flux ratios of
(a) [O {\sc ii}]$\lambda$3727/[N {\sc ii}]$\lambda$6583 versus
    [O {\sc i}]$\lambda$6300/[N {\sc ii}]$\lambda$6583, and
(b) [S {\sc ii}]$\lambda \lambda$6717,6731/[N {\sc ii}]$\lambda$6583 versus
    [O {\sc i}]$\lambda$6300/[N {\sc ii}]$\lambda$6583.
The asterisks, pluses, and ``Y'' signs denote the S1s, the S1.5s, and
the S2s of the sample of Nagao et al. (2001c), respectively.
The data are not corrected for the dust extinction.
The data points will move on the diagrams as shown by the arrow if the
extinction correction of $A_V$ = 1.0 mag is applied.
Note that the effect of the extinction correction in the diagram (b) 
is negligibly small.
\label{fig5}}
\end{figure*}

\section{NEW DIAGNOSTICS FOR THE GAS METALLICITY OF NLR}

 \subsection{Correlations among Forbidden Emission-Line Flux Ratios}

As mentioned in the last section, it seems desirable to use only forbidden
lines for studies of the gas metallicity of NLRs in S1s.
Furthermore, it seems ideal to use only low-ionization forbidden
emission lines to discuss the gas properties.
This is because low-ionization line-emitting regions and high-ionization
line-emitting regions may be spatially segregated (e.g., Baker 1997;
Murayama \& Taniguchi 1998a, 1998b; Nagao, Taniguchi, \& Murayama 2000;
Nagao, Murayama, \& Taniguchi 2001b, 2001c; see also
Hes, Barthel, \& Fosbury 1993).
Taking the above two constraints into account, we examine whether or not
the forbidden emission-line flux ratios of 
[O {\sc i}]$\lambda$6300/[N {\sc ii}]$\lambda$6583,
[O {\sc ii}]$\lambda$3727/[N {\sc ii}]$\lambda$6583, and
[S {\sc ii}]$\lambda \lambda$6717,6731/[N {\sc ii}]$\lambda$6583
can be useful estimators for the nitrogen abundance of gas clouds
in the NLRs.

In Figure 5, we show the relationship among the three forbidden emission-line
flux ratios. Here we use the data of Nagao et al. (2001c) in order
to find possible correlations among the three flux ratios.
Since the data are not corrected for the dust reddening, we also show 
the effect of the reddening correction, which is calculated by adopting 
the extinction curve of Cardelli et al. (1989), in Figure 5.
As shown apparently, positive correlations are seen among
the three flux ratios.
The corresponding correlation coefficients are 0.674 and 0.561 for
[O {\sc ii}]$\lambda$3727/[N {\sc ii}]$\lambda$6583 versus 
[O {\sc i}]$\lambda$6300/[N {\sc ii}]$\lambda$6583 and
[S {\sc ii}]$\lambda \lambda$6717,6731/[N {\sc ii}]$\lambda$6583 versus 
[O {\sc i}]$\lambda$6300/[N {\sc ii}]$\lambda$6583, respectively.
These correlation coefficients imply that there are
meaningful correlations among the three flux ratios.
What makes these correlations? Or, can these correlations be interpreted 
as sequences of the nitrogen abundance?
In order to investigate these issues, we perform photoionization model
calculations. The model methods and the results are presented in the
following sections.

 \subsection{Photoionization Models}

\begin{figure*}
\epsscale{1.00}
\plotone{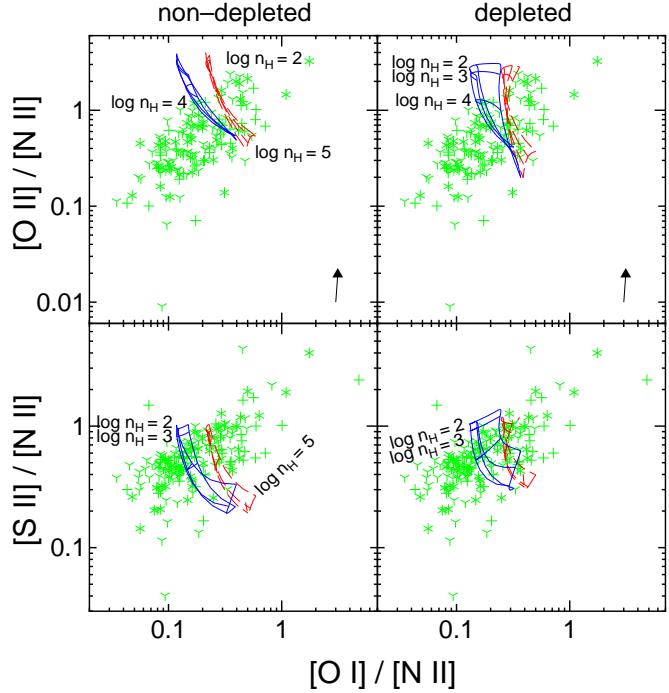}
\caption{
Diagnostic diagrams in which the data of Nagao et al. (2001c) are compared 
with the results of photoionization models. The symbols (green) and the 
arrow are the same as those in Figure 5. The red and the blue lines denote 
the predictions of the photoionization models with the NLS1-like SED and 
the BLS1-like SED, respectively. The results of the photoionization models 
with the non-depleted abundances and the depleted abundances are plotted in 
the left and the right panels, respectively.
\label{fig6}}
\end{figure*}

To investigate the origin of the correlations presented in the last section,
we carry out photoionization model calculations by using the publicly 
available code $Cloudy$ version 94.00 (Ferland 1997, 2000).
Here we assume uniform density gas clouds with a plane-parallel geometry.
The parameters for the calculations are 
(I)   the hydrogen density of a cloud ($n_{\rm H}$),
(II)  the ionization parameter ($U$),
(III) the chemical composition of the gas, and
(IV)  the shape of the spectral energy distribution (SED) of the 
      input continuum radiation.

We perform several model runs covering the following ranges of parameters:
10$^{2.0}$ cm$^{-3}$ $\leq$ $n_{\rm H}$ $\leq$ 10$^{5.0}$ cm$^{-3}$ and
10$^{-3.5}$ $\leq$ $U$ $\leq$ 10$^{-2.0}$.

For the chemical composition of the gas clouds,
we assume the case that the metals are all scaled keeping solar proportions
except for nitrogen; the nitrogen abundance scales with $Z^2$
(this assumption is altered in some calculations presented in \S 5).
We calculate the models covering a metallicity range of
0.25 $\leq$ $Z/Z_{\odot}$ $\leq$ 3.0,
which corresponds to 8.27 $\leq$ 12 + log (O/H) $\leq$ 9.35 and 
6.77 $\leq$ 12 + log (N/H) $\leq$ 8.92.
The adopted elemental abundances of the solar ones 
relative to hydrogen are taken from 
Grevesse \& Anders (1989) with extensions by Grevesse \& Noels (1993).
We also examine the case that a part of heavy elements is depleted into 
dust grains. In this case, 90\% of Mg, Si, and Fe, 50\% of C and O, and
25\% of N and S are locked into dust grains, as estimated for the Orion 
H {\sc ii} region (e.g., Baldwin et al. 1991, 1996). This corresponds to 
the case that the gas clouds in NLRs contain unnegligible dust grains.
However, we ignore grain opacity, heating and cooling, since their properties
in photoionized gases are highly uncertain. Although the heating process by
grain photoelectrons is important for enhancing some high-ionization emission
lines (e.g., Shields \& Kennicutt 1995), it does not affect our results
significantly since we use only low-ionization emission lines.

We adopt two types of SED for the calculations; for both the SEDs, 
we adopt the following function:
\begin{equation}
f_{\nu} = 
 \nu^{\alpha_{{\rm uv}}} \exp \left( -\frac{h\nu}{kT_{{\rm BB}}} \right) 
 \exp \left( -\frac{kT_{{\rm IR}}}{h\nu} \right) + a\nu^{\alpha_{{\rm x}}}
\end{equation}
(see Ferland 1997).
We adopt the following parameter set (see Nagao et al. 2001a for more 
details):
(i)   the infrared cutoff of the big blue bump component, $kT_{\rm IR}$ = 
      0.01 Ryd,
(ii)  the slope of the low-energy side of the big blue bump, 
      $\alpha_{\rm uv}$ = --0.5,
(iii) the UV--to--X-ray spectral index, $\alpha_{\rm ox}$ = --1.35,
(iv)  the slope of the X-ray power-law continuum, $\alpha_{\rm x}$ = 
      --0.85 or --1.15, and
(v)   the characteristic temperature of the big blue bump, $T_{\rm BB}$
      = 4.9 $\times$ 10$^5$ K or 11.8 $\times$ 10$^5$ K.
Hereafter the continua with ($\alpha_{\rm x}$, $T_{\rm BB}$) = 
(--0.85, 4.9 $\times$ 10$^5$ K) and (--1.15, 11.8 $\times$ 10$^5$ K) 
are denoted as ``BLS1-like SED'' and ``NLS1-like SED'', respectively.
Note that the parameter $a$ in equation (3) is determined from
the adopted value of $\alpha_{\rm ox}$.
The last term in equation (3) is not extrapolated below 1.36 eV or
above 100 keV. Below 1.36 eV the last term is simply set to zero.
Above 100 keV the continuum is assumed to fall off as $\nu^{-3}$.

The calculations are stopped when the temperature falls to 3000 K, below
which the gas does not contribute significantly to the observed
optical emission-line spectra.

 \subsection{Results of Model Calculations}

\begin{figure*}
\epsscale{1.00}
\plotone{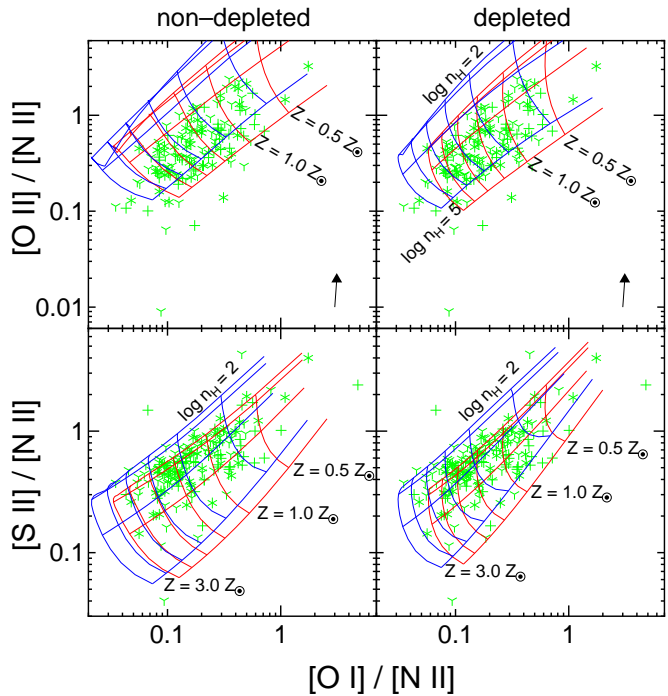}
\caption{
Same as Figure 6, but other model grids are overdrawn. As for the 
photoionization models, the ionization parameter is fixed to be 
$U = 10^{-3.0}$. The metallicity is varied between 0.25 $\leq$ $Z/Z_{\odot}$ 
$\leq$ 3.0. The symbols, the lines, and the arrow are the same as those 
in Figure 6.
\label{fig7}}
\end{figure*}

In Figure 6, we present the results of the models with the metallicity of
$Z = Z_{\odot}$.
These model results are used to
investigate whether or not the gas density, the ionization parameter, or
the combination of both two parameters drives the correlations presented in
\S 4.1. Here the models with the NLS1-like SED and the BLS1-like 
SED are presented.
It is apparently shown that the emission-line flux ratios of 
[O {\sc i}]$\lambda$6300/[N {\sc ii}]$\lambda$6583,
[O {\sc ii}]$\lambda$3727/[N {\sc ii}]$\lambda$6583, and
[S {\sc ii}]$\lambda \lambda$6717,6731/[N {\sc ii}]$\lambda$6583
are insensitive to the ionization parameter for any type of the
adopted ionizing SEDs. The predicted emission-line flux ratios vary 
by less than a factor of 2 when
the ionization parameter is changed between 10$^{-3.5}$ and 10$^{-2.0}$.
This is mainly because we use only low-ionization forbidden emission lines.
Although the three emission-line flux ratios are rather sensitive to the 
gas density, the model grids for varying gas densities are perpendicular to
the observed correlations for any type of the adopted ionizing SEDs, 
as shown in Figure 6.
These results do not depend on whether or not the heavy elements are 
depleted into dust grains, as shown in Figure 6.
Therefore, we conclude that the observed correlations among the 
three emission-line flux ratios are not significantly attributed to 
the effects of the gas density, the ionization parameter, or both.

Then, are the correlations 
attributed to the nitrogen abundance (i.e., the gas metallicity) ? 
In Figure 7, we show the results of model 
calculations in which the gas metallicity is varied between 
$0.25 \leq Z/Z_{\odot} \leq 3.0$.
Here the NLS1-like SED and the BLS1-like SED are adopted for the ionizing
continuum together with $U = 10^{-3.0}$.
It is shown that the metallicity sequences can well explain
the observed correlations in the flux ratios of the low-ionization 
forbidden lines for both cases, the models with the non-depleted 
abundances and with the depleted abundances.
We thus conclude that the observed correlations among the 
flux ratios of the low-ionization forbidden lines are mainly attributed to
the variety of
the gas metallicity. Since the narrow component of permitted lines is
not used and only forbidden lines are used, these diagrams presented in
Figure 7 can be powerful diagnostics for the gas metallicity in the NLR of
AGNs, especially for S1s.

Here we mention that the flux ratio of 
[O {\sc i}]$\lambda$6300/[N {\sc ii}]$\lambda$6583 seems to be the
most appropriate diagnostic for gas metallicity among the three
emission-line flux ratios. This is because the flux ratio of
[O {\sc ii}]$\lambda$3727/[N {\sc ii}]$\lambda$6583 and
[S {\sc ii}]$\lambda \lambda$6717,6731/[N {\sc ii}]$\lambda$6583
are sensitive to the gas density while that of 
[O {\sc i}]$\lambda$6300/[N {\sc ii}]$\lambda$6583 is not, unless the 
gas density is high ($\gtrsim 10^4$ cm$^{-3}$).
The flux ratio of [O {\sc ii}]$\lambda$3727/[N {\sc ii}]$\lambda$6583
is sensitive to the dust extinction, additionally.

 \subsection{Application to NLRs in NLS1s and BLS1s}

\begin{figure*}
\epsscale{1.00}
\plotone{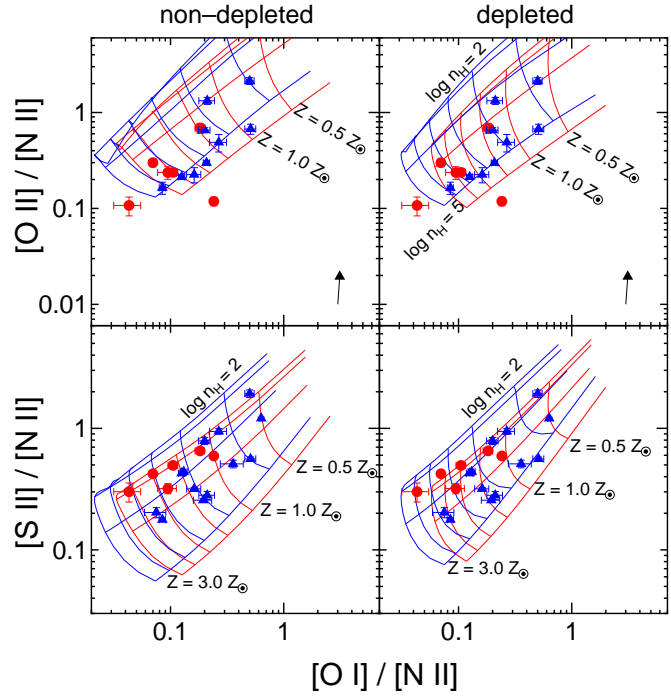}
\caption{
Same as Figure 7, but the data of Rodr\'{\i}guez-Ardila et al. (2000)
are plotted instead of the data of Nagao et al. (2001c).
The lines and the arrow are the same as those in Figure 6.
The red circles and the blue triangles denote the data of the NLS1s and 
the BLS1s in the sample of Rodr\'{\i}guez-Ardila et al. (2000),
respectively.
\label{fig8}}
\end{figure*}

Now we apply our new method to the investigation of the gas metallicity
of NLRs in NLS1s and BLS1s.
In Figure 8, we plot the data of Rodr\'{\i}guez-Ardila et al. (2000) on the 
diagnostic diagrams presented in \S 4.3.
It appears that the NLS1s have smaller flux ratios of
[O {\sc i}]$\lambda$6300/[N {\sc ii}]$\lambda$6583,
[O {\sc ii}]$\lambda$3727/[N {\sc ii}]$\lambda$6583, and
[S {\sc ii}]$\lambda \lambda$6717,6731/[N {\sc ii}]$\lambda$6583
than the BLS1s. The average and median values of these three emission-line 
flux ratios for the NLS1s and BLS1s are given in Table 1.
These values imply that there are 
differences in the emission-line flux 
ratios between the NLS1s and the BLS1s, though the statistical significance 
is low; i.e., $P_{\rm KS} =$ 0.133, 0.533, and 0.767 for the flux ratios of
[O {\sc i}]$\lambda$6300/[N {\sc ii}]$\lambda$6583,
[O {\sc ii}]$\lambda$3727/[N {\sc ii}]$\lambda$6583, and
[S {\sc ii}]$\lambda \lambda$6717,6731/[N {\sc ii}]$\lambda$6583,
respectively (see Table 1). Here it should be noted that 
the photoionization models predict larger ratios of
[O {\sc i}]$\lambda$6300/[N {\sc ii}]$\lambda$6583 for NLS1s
than for BLS1s, if the difference in the SED of ionizing
photons between NLS1s and BLS1s is taken into account.
The data of Rodr\'{\i}guez-Ardila et al. (2000) exhibit the opposite trend
from the prediction of the photoionization models.
It is thus suggested that the gas metallicity in NLRs of the NLS1s
is possibly higher than that of the BLS1s, at least for the sample of
Rodr\'{\i}guez-Ardila et al. (2000).

In Figure 9, we show the relationship between the FWHM of the broad 
component of the H$\alpha$ emission and the flux ratio of
[O {\sc i}]$\lambda$6300/[N {\sc ii}]$\lambda$6583 for the sample of
Rodr\'{\i}guez-Ardila et al. (2000). Being different from Figures 2 and 3,
it appears to be a positive correlation between them.
Its correlation coefficient is 0.681, which implies that
there is a meaningful correlation between them.
This is consistent with the result of Wills et al. (1999); that is,
the object with narrower BLR emission tends to exhibit
higher nitrogen abundance.

\section{DISCUSSION}

\begin{figure*}
\epsscale{1.00}
\plotone{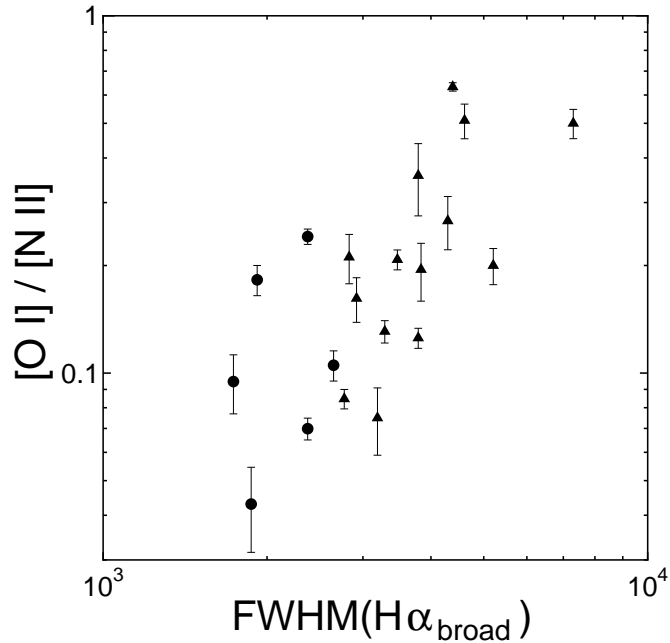}
\caption{
Diagram of the flux ratio of 
[O {\sc i}]$\lambda$6300/[N {\sc ii}]$\lambda$6583 versus
the FWHM of the broad component of the H$\alpha$ emission.
The symbols are the same as those in Figure 2.
\label{fig9}}
\end{figure*}

As investigated in \S 3 and \S 4, the inferred difference in the 
gas metallicity of NLRs between NLS1s and BLS1s depends on the
adopted diagnostics.
When equation (1) is used to estimate the gas metallicity,
it is suggested that NLS1s tend to have metal-poorer NLR gas
than BLS1s, though the difference is not significant.
Contrary to this result, the diagnostics examined in
\S 4 suggest that NLS1s appear to possess metal-richer NLR gas
than BLS1s. There appears to be no significant difference in the
estimated metallicity by using equation (2).
Why the results are inconsistent in each other?
Which is a reliable diagnostics to estimate the gas metallicity
in NLRs?
Here we discuss these issues in order to understand how the gas
metallicity is different, or not, between NLS1s and BLS1s.

As discussed by Nagao et al. (2001c), it is sometimes difficult to
measure fluxes of the narrow component of permitted lines for S1s.
Since the method using either equation (1) or (2) requires the flux
of the narrow component of permitted lines, the oxygen abundances
derived at \S 3 can suffer from this effect
and thus he derived values may contain large errors.
In addition to this uncertainty, there is another effect which may
diminish the reliability of the metallicity diagnostics adopted in 
\S 3. V\'{e}ron-Cetty et al. (2001) reported that
the broad Balmer lines in the spectra of NLS1s are well fitted by
a Lorentzian profile rather than a Gaussian profile
(see also Moran et al. 1996; Leighly 1999b; Gon\c calves et al. 1999;
Sulentic et al. 2002).
When the broad component is fitted by a Lorentzian profile instead of a
Gaussian profile, the measurement of the narrow component of the line is
affected. This results in larger ratios of
[N {\sc ii}]$\lambda$6583/H$\alpha_{\rm narrow}$ and 
[O {\sc iii}]$\lambda$5007/H$\beta_{\rm narrow}$
compared to the case when the broad component of permitted lines is fitted
by a Gaussian profile as done by Rodr\'{\i}guez-Ardila et al. (2000)
(see Figure 8 of V\'{e}ron-Cetty et al. 2001).
Therefore, the ratios of [N {\sc ii}]$\lambda$6583/H$\alpha_{\rm narrow}$ 
and [O {\sc iii}]$\lambda$5007/H$\beta_{\rm narrow}$ in the data of 
Rodr\'{\i}guez-Ardila et al. (2000) may be underestimated if
a Lorentzian profile is a better function to measure the broad component 
of permitted lines than a Gaussian profile which 
Rodr\'{\i}guez-Ardila et al. (2000) adopted. 
In order to see how the underestimation of the flux ratios of 
[N {\sc ii}]$\lambda$6583/H$\alpha_{\rm narrow}$ and 
[O {\sc iii}]$\lambda$5007/H$\beta_{\rm narrow}$ affects the inferred oxygen 
abundance, we consider the case in which the two flux ratios are 
underestimated by a factor of 10.
We assume [N {\sc ii}]$\lambda$6583/H$\alpha_{\rm narrow}$ = 2.0,
[O {\sc iii}]$\lambda$5007/H$\beta_{\rm narrow}$ = 10.0, and
[O {\sc ii}]$\lambda$3727/[O {\sc iii}]$\lambda$5007 = 0.1, 
which would be measured as
[N {\sc ii}]$\lambda$6583/H$\alpha_{\rm narrow}$ = 0.2,
[O {\sc iii}]$\lambda$5007/H$\beta_{\rm narrow}$ = 1.0, and
[O {\sc ii}]$\lambda$3727/[O {\sc iii}]$\lambda$5007 = 0.1, 
respectively. In this situation, the oxygen abundances are also 
underestimated by a factor of $\sim$0.28 for log (O/H)$_{\rm SSCK1}$
and by a factor of $\sim$0.30 for log (O/H)$_{\rm SSCK2}$.
This result suggests that we might underestimate the oxygen abundance
in \S 3 if the Lorentzian profile is the representative function for
the broad component of permitted lines. 
This may be the reason why the results derived in \S 3 and \S 4 are
inconsistent; the oxygen abundance of gas in NLS1s may be selectively
underestimated when the SSCK method is applied, due to the 
Lorentzian profile of broad components of permitted lines
in the spectra of NLS1s.
Although the reason why the broad component of the permitted lines is
well fitted by Lorentzian is not clear, Dumont \& Collin-Souffrin (1990)
showed that the Lorentzian profile could be produced if the BLR emission is
produced at gas clouds in a disk-like geometry around a supermassive
black hole (see also V\'{e}ron-Cetty et al. 2001; Sulentic et al. 2002).
We do not discuss the origin of the Lorentzian profile 
further since this issue is beyond the purpose of this paper.
Since the diagnostics investigated in \S 4 consist of only forbidden 
lines and thus are free from the above uncertainties, we
conclude that the results obtained in \S 4 are more reliable
than those obtained in \S 3. Therefore, it can be concluded that
the metallicity of gas clouds in NLRs is higher in NLS1s than in BLS1s.

\begin{figure*}
\epsscale{1.00}
\plotone{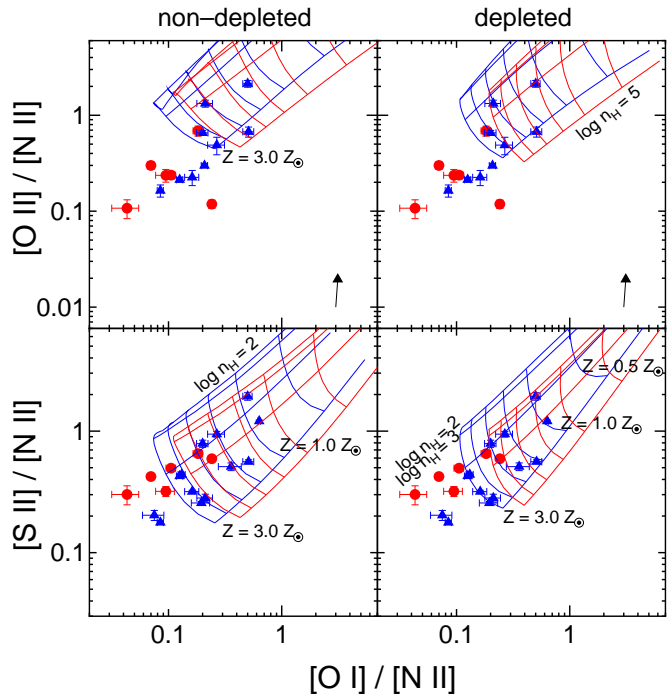}
\caption{
Same as Figure 9, but the models with $q = 0.5$ are drawn
instead of the models with $q = 0$.
\label{fig10}}
\end{figure*}

Here we mention that the diagnostics investigated in \S 4 may
also have a possible drawback.
As shown in Figure 7, the correlations among the emission-line flux ratios of
[O {\sc i}]$\lambda$6300/[N {\sc ii}]$\lambda$6583,
[O {\sc ii}]$\lambda$3727/[N {\sc ii}]$\lambda$6583, and
[S {\sc ii}]$\lambda \lambda$6717,6731/[N {\sc ii}]$\lambda$6583 can be
explained by the photoionization models in the parameter ranges of
$n_{\rm H} \sim 10^{2-5}$ cm$^{-3}$ and 
0.5 $\lesssim Z/Z_{\odot} \lesssim$ 3.0.
The inferred range of the gas density is consistent with the previously
reported values (e.g., Tsvetanov \& Walsh 1992;
Storchi-Bergmann, Wilson, \& Baldwin 1992; Schmitt et al. 1994).
However, the model values of metallicity
would contain possible uncertainties if the nitrogen production is 
delayed relative to the oxygen production.
The delay is significant if the timescale for the increase in the 
metallicity by star formation activity is comparable to or shorter than 
the lifetime of the nitrogen-producing stars, i.e., $\sim 10^8$ yr
(e.g., Hamann \& Ferland 1993, 1999; 
Henry, Edmunds, \& K\"{o}ppen 2000; see also Hamann et al. 2002).
This effect results in that the abundance ratio of N/H become to be below the
solar value at $Z = Z_{\odot}$.
Although this offset is only weakly present in the data of H {\sc ii}
region (van Zee et al. 1998), it could be substantial in AGNs
(see, e.g., Hamann \& Ferland 1999).
Taking this effect into account, one can formulate the nitrogen abundance 
of the gas as
\begin{equation}
{\rm log} \frac{\rm (N/H)}{\rm (N/H)_{\odot}} = 2 {\rm log} (Z/Z_{\odot}) - q
\end{equation}
where $q$ is the logarithmic offset.
Hamann et al. (2002) mentioned that the $q$ value might reach up to
$\sim$0.2--0.5 for gas in AGNs.
In order to investigate the effect of this delay on our diagnostics,
we perform further model calculations. The adopted parameters are the same as 
those of the models presented in Figures 7 and 8
except for the $q$ value, which is altered from
$q = 0$ to $q = 0.5$. The results are shown in Figure 10.
The diagrams in Figure 10 suggest that the gas metallicity of NLRs is
$1.0 \lesssim Z/Z_{\odot} \lesssim 3.0$ for the BLS1s while
$Z/Z_{\odot} \gtrsim 2.5$ for the NLS1s.
These values are larger than those suggested by the diagrams in Figure 8.
Therefore, the diagnostics investigated in \S 4 may be invalid
when we estimate the absolute values of the gas metallicity
since the true $q$ value for the NLR in AGNs is unclear.
We can, however, use the diagnostics to compare the gas metallicities
between NLS1s and BLS1s unless the $q$ value is systematically different
between NLS1s and BLS1s.
Since there is no reason to exist such a difference in 
the $q$ value between the two populations of S1s, the conclusion that 
the gas metallicity in NLRs tends to
be relatively higher in NLS1s than in BLS1s seems valid, 
even if the uncertainty of the $q$ value is taken into account.

Finally, we discuss the effect of the inhomogeneity of NLRs on our 
results. It is known that there are significant density gradients in NLRs 
(e.g., Kraemer et al. 2000). Since the [O {\sc i}]$\lambda$6300 transition 
has higher critical density ($1.8 \times 10^6$ cm$^{-3}$) than the transitions 
of [O {\sc ii}]$\lambda$3727, [N {\sc ii}]$\lambda$6583, 
[S {\sc ii}]$\lambda$6717, and [S {\sc ii}]$\lambda$6731 
($4.5 \times 10^3$ cm$^{-3}$, $8.7 \times 10^4$ cm$^{-3}$, 
$1.5 \times 10^3$ cm$^{-3}$, and $3.9 \times 10^3$ cm$^{-3}$, respectively), 
gas clouds with $n_{\rm H} > 10^5$ cm$^{-3}$ may selectively radiate the 
[O {\sc i}]$\lambda$6300 emission. 
Therefore our one-zone photoionization models may be inappropriate to 
investigate the gas metallicity of NLRs if such a high-density gas cloud 
contributes to emission-line spectra of NLS1s and BLS1s differently. Indeed 
there are significant differences in the emission-line fluxes of some 
high-ionization emission line such as [Fe {\sc vii}]$\lambda$6087 and 
[Fe {\sc x}]$\lambda$6374 between S1s and S2s, which is due to the 
different contribution of dense gas clouds in NLRs (e.g., 
Murayama \& Taniguchi 1998a; Nagao et al. 2000, 2001c). However, it has 
been already shown by Nagao et al. (2000, 2001c) that there is no 
systematic difference in the high-ionization emission lines between NLS1s 
and BLS1s. This suggests that the contribution from the dense gas clouds to 
emission-line spectra is not so different between NLS1s and BLS1s. 
Therefore, our conclusions seem valid even if possible density 
inhomogeneity is taken into account.

\section{SUMMARY}

In order to explore possible differences in the gas metallicity between
NLS1s and BLS1s, we have focused on the gas in NLRs.
It is found that the results depend on the applied diagnostics.
The emission-line flux ratio of 
[N {\sc ii}]$\lambda$6538/H$\alpha_{\rm narrow}$, which has been often
used to discuss the gas metallicity of NLRs in S2s, is not significantly
different between NLS1s and BLS1s.
Some other diagnostics in which the narrow component of permitted lines is
used also suggest that there is little or no systematic difference
in the gas metallicity between NLS1s and BLS1s. 
On the other hand, the diagnostics which consist of only forbidden 
emission lines, which are newly proposed in this paper, suggest that
the gas metallicity of NLRs in NLS1s tend to be higher than that in BLS1s.
This complex situation may be solved if the Lorentzian profile, rather than
the Gaussian profile, is the representative function for
the broad component of permitted lines in the spectra of NLS1s.
Since the results derived by using only forbidden emission-line fluxes
do not suffer from the uncertainty of the fitting function for the broad 
component of permitted lines, it can be concluded that the gas metallicity 
of NLRs tends to be higher in NLS1s than in BLS1s.

\acknowledgments

We would like to thank Gary Ferland for providing his code $Cloudy$ 
to the public. Alberto Rodr\'{\i}guez-Ardila kindly provided information 
about the emission-line flux ratios of some Seyfert galaxies.
YS is supported by a Research Fellowship from the Japan 
Society for the Promotion of Science for Young Scientists.
This work was financially supported in part by Grant-in-Aids for the 
Scientific Research (Nos. 10044052, 10304013, and 13740122) of the 
Japanese Ministry of Education, Culture, Sports, Science, and Technology.


\end{document}